# Profitable Scheduling on Multiple Speed-Scalable Processors[*]

Peter Kling and Peter Pietrzyk

**Heinz Nixdorf Institute and Computer Science Department**
**University of Paderborn**
**Fürstenallee 11, 33102 Paderborn, Germany**

## Abstract

We present a new online algorithm for profit-oriented scheduling on multiple speed-scalable processors. Moreover, we provide a tight analysis of the algorithm's competitiveness. Our results generalize and improve upon work by Chan, Lam, and Li [10], which considers a single speed-scalable processor. Using significantly different techniques, we can not only extend their model to multiprocessors but also prove an enhanced and tight competitive ratio for our algorithm.

In our scheduling problem, jobs arrive over time and are preemptable. They have different workloads, values, and deadlines. The scheduler may decide not to finish a job but instead to suffer a loss equaling the job's value. However, to process a job's workload until its deadline the scheduler must invest a certain amount of energy. The cost of a schedule is the sum of lost values and invested energy. In order to finish a job the scheduler has to determine which processors to use and set their speeds accordingly. A processor's energy consumption is power $P_\alpha(s)$ integrated over time, where $P_\alpha(s) = s^\alpha$ is the power consumption when running at speed $s$. Since we consider the online variant of the problem, the scheduler has no knowledge about future jobs. This problem was introduced by Chan, Lam, and Li [10] for the case of a single processor. They presented an online algorithm which is $\alpha^\alpha + 2e\alpha$-competitive. We provide an online algorithm for the case of multiple processors with an improved competitive ratio of $\alpha^\alpha$.

## 1 Introduction

From an economical point of view, the value of energy has increased tremendously during the last decades. This applies not only to the energy consumed in small-scale computer systems but especially to the energy consumption in large data centers. According to current reports (e.g., Barroso and Hölzle [6]), the decisive factors regarding the costs of running a data center are mostly the cooling process and the actual computations rather than the acquisition of the necessary hardware. Thus, in order to maximize their revenue, data centers strive to minimize the energy consumption while still guaranteeing a sufficiently high quality of service to their customers. One way to approach this goal are technical solutions improving the involved hardware. However, coupling such solutions with canonical or standard algorithms wastes much potential. Only by designing sophisticated algorithms can one hope to fully exploit their power and possibilities. A prominent example for this is *dynamic speed scaling*, a technology that adapts a processor's speed according to the current workload (*Intel SpeedStep* or *AMD PowerNow!*). Simply decreasing the speed at times of small load may lower the total energy consumption substantially. However, a lower speed often also implies a lower quality of service, which in turn may impair the data center's revenue. One needs clever algorithms to fully utilize speed scaling and to achieve a provably good or even optimal profit.

* This work is partially supported by the German Research Foundation (DFG) within the Collaborative Research Center "On-The-Fly Computing" (SFB 901) and by the Graduate School on Applied Network Science (GSANS).

But how exactly should a data center make use of speed scaling in order to maximize profit? On a relatively basic level, one can imagine a data center's situation as follows: Jobs of different sizes and values arrive over time at the data center. For finishing a customer's job in time, the data center receives a payment corresponding to the job's value. However, to finish a job the data center has to invest an amount of energy depending on the job's size and potential time constraints. Investing into low-value jobs that require much energy may lower the profit. Even processing jobs whose values seem to justify the energy investment may be bad, as this may hinder the efficient processing of more lucrative jobs that arrive later. Thus, one has to carefully choose not only how and when to process the different jobs but also which to process at all. We propose an algorithm that handles this scenario provably well and improves upon the former best known result. Moreover, we generalize the model to the important case of multiple processors (until now, only a single speed-scalable processor was considered). Our analysis is partly based on an intriguing new technique recently suggested by Gupta, Krishnaswamy, and Pruhs [12]. We adapt and extend it to suit our problem and show its large potential compared to the classical analysis methods prevailing in this area (see "Our Contribution" later in this section).

**Related Work.**

There exists plenty of work concerning energy-efficient scheduling strategies in both theoretical and practical contexts. Dynamic speed scaling (also referred to as dynamic voltage scaling) is one of the most important technical tools to save energy in modern systems. It allows the scheduler to dynamically adapt the system's speed to the current workload. A recent survey by Albers [1] gives a good and compact overview on the state of the art of algorithmic research in this area. In the following, we concentrate on models for speed-scalable processors and jobs with deadline constraints. Theoretical work in this area has been initiated by Yao, Demers, and Shenker [14]. They considered a single speed-scalable processor that processes preemptable jobs which arrive over time and come with different deadlines and workloads. Yao, Demers, and Shenker studied the question of how to finish all the jobs in an energy-minimal way. In their seminal work [14], they modeled the power consumption $P_\alpha(s)$ of a processor running at speed $s$ by a constant degree polynomial $P_\alpha(s) = s^\alpha$. Here, the energy exponent $\alpha$ is assumed to be a constant $\alpha \geq 2$. In classical CMOS-based systems $\alpha = 3$ usually yields a suitable approximation of the actual power consumption. Yao, Demers, and Shenker developed an optimal offline algorithm, known as YDS, as well as the two online algorithms *Optimal Available* (OA) and *Average Rate* (AVR). Up to now, OA remains one of the most important algorithms in this area, being an essential part of many algorithms for both the original problem as well as for its manifold variations. Using a rather complex but elegant amortized potential function argument, Bansal, Kimbrel, and Pruhs [3] proved that OA is exactly $\alpha^\alpha$-competitive. They also proposed a new algorithm, named BKP, which achieves a competitive ratio of essentially $2e^{\alpha+1}$. The algorithm qOA presented by Bansal et al. [5] is particularly well suited for low powers of $\alpha$, where it outperforms both OA and BKP. In this work, the authors also proved that no deterministic algorithm can achieve a competitive ratio of better than $e^{\alpha-1}/\alpha$. In their recent work, Albers, Antoniadis, and Greiner [2] presented an optimal offline algorithm for the multiprocessor case. Moreover, using this algorithm, they were able to also extend OA to the multiprocessor case and proved the same competitive ratio of $\alpha^\alpha$ as in the single processor case.

All results mentioned so far are concerned only with the energy necessary to finish *all* jobs. With respect to the profitability aspect, the two most relevant results for us are due to Chan, Lam, and Li [10] and Pruhs and Stein [13]. Both proposed a model incorporating profitability into classical energy-efficient scheduling. In the simplest case, jobs have values and the scheduler is no longer required to finish all jobs. Instead, it can decide to not process jobs whose values do not justify the foreseeable energy investment necessary to complete them. The objective is to maximize the profit [13] or, similarly, to minimize the loss [10]. As argued by the authors, the latter model has the benefit of being a direct generalization of the classical model by Yao, Demers, and Shenker. For maximizing the profit, Pruhs and Stein [13] showed that in order to achieve a bounded competitive ratio, resource augmentation is necessary and gave a scalable online algorithm. For minimizing the loss, Chan, Lam, and Li [10] gave an $\alpha^\alpha + 2e\alpha$-competitive algorithm. Another very important and recent work is due to Gupta, Krishnaswamy, and Pruhs [12] and considers the *Online Generalized Assignment Problem* (ONGAP). The authors showed an interesting relation to a multitude of problems in the context of speed-scalability (not only for scheduling). They developed a convex programming formulation of the problem and applied well-known techniques from convex optimization. Especially, they used a greedy primal-dual approach as known from linear programming (see, e.g., [9]). This way, they designed an online algorithm for the classical model by Yao, Demers, and Shenker (no job values; one processor) which is very similar to OA and proved the exact same competitive ratio of $\alpha^\alpha$.

**Our Contribution.**

We develop and analyze a new online algorithm for scheduling valuable jobs on multiple speed-scalable processors. Our algorithm improves upon known results in two respects: For the single processor case it improves the best known competitive ratio from $\alpha^\alpha + 2e\alpha$ to $\alpha^\alpha$. Moreover, this constant competitive ratio holds even for the case of multiple processors. To the best of our knowledge, this is the first algorithm that is able to handle the multiprocessor case in this scenario. We also show that our analysis is tight in that the proven competitive ratio is optimal for our algorithm.

Our analysis is significantly different from the typical potential function argument which is dominant in the analysis of online algorithms in this research area. Instead, we make use of a framework recently suggested by Gupta, Krishnaswamy, and Pruhs [12]. It utilizes well-known tools from convex optimization, especially duality theory and primal-dual algorithms. We develop a convex programming formulation and design a greedy primal-dual online algorithm for the problem at hand. Compared to the original framework, we have to overcome the additional issue of integral variables in our convex program that are caused by the new profitability aspect. Moreover, the handling of multiple processors proves to be a challenging task. It not only causes a much more complex objective function in the convex program but also makes it harder to grasp the structural properties of the resulting schedule. Our result shows that this technique is not only suitable for the classical energy-efficient scheduling model but also for more complex variations of it, as conjectured by Gupta, Krishnaswamy, and Pruhs. It is interesting to note that, in terms of the analysis, this approach goes back to the roots of Yao, Demers, and Shenker's model, as the optimality proof of the YDS algorithm [4] is based on a similar convex programming formulation and the well-known KKT conditions from convex optimization [8]. Our algorithm can be seen as greedily increasing the convex program's variables while maintaining a relaxed version of these KKT conditions.

## 2   Model & Preliminaries

We consider a system of $m$ speed-scalable processors. That is, each processor can be set to any speed $s \in \mathbb{R}_{\geq 0}$ (independently from the others). When running at speed $s$, the power consumption of a single processor is given by the *power function* $P_\alpha(s) = s^\alpha$. Here, the constant parameter $\alpha \in \mathbb{R}_{>1}$ is called the *energy exponent*. A problem instance consists of a set $J = \{1, 2, \ldots, n\}$ of $n$ jobs. Each job $j \in J$ is associated with a *release time* $r_j$, a *deadline* $d_j$, a *workload* $w_j$, and a *value* $v_j$. A *schedule* $S$ describes if and how the different jobs are processed by the system. It consists of $m$ speed functions $S_i \colon \mathbb{R}_{\geq 0} \to \mathbb{R}_{\geq 0}$ ($i \in \{1, 2, \ldots, m\}$) and a job assignment policy. The speed function $S_i$ dictates the speed $S_i(t)$ of the $i$-th processor at time $t$. The job assignment policy decides which jobs to run on the processors. At any time $t$, it may schedule at most one job per processor, and each job can be processed by at most one processor at any given time (i.e., we consider nonparallel jobs). Moreover, jobs are preemptive: a running job may be interrupted at any time and continued later on, possibly on a different processor. The total work processed by processor $i$ between time $t_1$ and $t_2$ is $\int_{t_1}^{t_2} S_i(t)\,\mathrm{d}t$. Similarly, the overall power consumed by this processor during the same time is $\int_{t_1}^{t_2} P_\alpha(S_i(t))\,\mathrm{d}t$. Let $s_j(t)$ denote the speed used to process job $j$ at time $t$. We say job $j$ is *finished under schedule* $S$ if $S$ processes (at least) $w_j$ units of $j$'s work during the interval $[r_j, d_j)$. That is, if we have $\int_{r_j}^{d_j} s_j(t)\,\mathrm{d}t \geq w_j$.

A given schedule $S$ may not finish all $n$ jobs. In this case, the total value of unfinished jobs is considered as a loss. Thus, the cost of $S$ is defined as the sum of the total energy consumption and the total value of unfinished jobs. More formally, if $J_{\mathrm{rej}}$ denotes the set of unfinished (aka rejected) jobs under schedule $S$, we define the *cost of schedule $S$* by

$$\mathrm{cost}(S) := \sum_{i=1}^{m} \int_0^\infty P_\alpha(S_i(t))\,\mathrm{d}t + \sum_{j \in J_{\mathrm{rej}}} v_j. \tag{1}$$

Our goal is to construct a low-cost schedule in the *online scenario* of the problem. That is, the job set $J$ is not known a priori, but rather revealed over time. Especially, we do not know the total number of jobs, and the existence as well as the attributes of a job $j \in J$ are revealed just when the job is released at time $r_j$. We measure the quality of algorithms for this online problem by their *competitive ratio*: Given an online algorithm $A$, let $A(J)$ denote the resulting schedule for job set $J$. The competitive ratio of $A$ is defined as

$$\sup_J \frac{\mathrm{cost}(A(J))}{\mathrm{cost}(\mathrm{OPT}(J))}, \tag{2}$$

where $\mathrm{OPT}(J)$ denotes an optimal schedule for the job set $J$. Note that, by definition, the competitive ratio is at least one.

## 2.1 Convex Programming Formulation

In the following, we develop a convex programming formulation of the above (offline) scheduling problem to aid us in the design and analysis of our online algorithm (cf. Section 3). Following an idea by Bingham and Greenstreet [7], we partition time into *atomic intervals* $T_k$ using the jobs' release times and deadlines. The goal of our convex program is to compute what portion of each job to process during the different atomic intervals in an optimal schedule. Once we have such a fixed *work assignment*, we use a deterministic algorithm by Chen et al. [11] to efficiently compute an energy-minimal way to process the corresponding work on the $m$ processors in this interval. The energy consumption of the resulting schedule in the interval $T_k$ can be written as a convex function $\mathcal{P}_k$ of the work assignment. This function plays a crucial role in the optimization objective of our convex program, and studying its properties and the corresponding schedule's structure is an important part of our analysis. We will elaborate on $\mathcal{P}_k$ once we have derived the convex program (see Section 2.2).

$$\min_{\substack{0 \preceq x \\ y \in \{0,1\}^n}} \quad \sum_{k=1}^{N} \mathcal{P}_k(x_{1k}, x_{2k}, \ldots, x_{nk}) + \sum_{j \in J}(1 - y_j)v_j$$

$$\text{s.t.} \quad y_j - \sum_{k=1}^{N} c_{jk}x_{jk} \leq 0 \qquad\qquad , j \in J$$

■ **Figure 1** Mathematical programming formulation (IMP) of our scheduling problem.

For a given job set $J$, let us partition the time horizon into $N \in \mathbb{N}$ atomic intervals $T_k$ ($k \in \{1, 2, \ldots, N\}$) as follows. We define $T_k := [\tau_{k-1}, \tau_k]$ where $\tau_0 < \tau_1 < \ldots < \tau_N$ are chosen such that $\{\tau_0, \tau_1, \ldots, \tau_N\} = \{r_j, d_j \mid j \in J\}$. Let $l_k := \tau_k - \tau_{k-1}$ denote the length of interval $T_k$. Note that there are at most $2n - 1$ intervals. To model the deadline constraint of job $j$, we introduce parameters $c_{jk} \in \{0, 1\}$ that indicate whether $T_k \subseteq [r_j, d_j)$ ($c_{jk} = 1$) or not ($c_{jk} = 0$). Our program uses two types of variables: *load variables* $x_{jk} \in [0, 1]$ for each job $j \in J$ and each atomic interval $k \in \{1, 2, \ldots, N\}$, and *indicator variables* $y_j \in \{0, 1\}$ for each job $j \in J$. The variable $x_{jk}$ indicates what portion of $j$'s workload is assigned to interval $T_k$ and the variable $y_j$ indicates whether job $j$ is finished ($y_j = 1$) or not ($y_j = 0$). Figure 1 shows the complete (integral) mathematical program (IMP) for our scheduling problem. The first summand in the objective corresponds to the energy spent in the different intervals. The second summand charges costs for all unfinished jobs. The set of constraints ensures that a job can be declared as finished only if it has been completely assigned to intervals $T_k$ lying in its release-deadline interval $[r_j, d_j)$. We use $x$ and $y$ to refer to the full vectors of variables $x_{jk}$ and $y_j$, and we use the symbol "$\preceq$" for element-wise comparison.

If we relax the domain of (IMP) such that $0 \preceq y \preceq 1$, we get a convex program. We refer to this convex program as (CP). By introducing dual variables $\lambda_j$ (also called *Lagrange multipliers*) for each constraint of (CP) we can write its *Lagrangian* $L(x, y, \lambda)$ as

$$\sum_{k=1}^{N} \mathcal{P}_k(x_{1k}, x_{2k}, \ldots, x_{nk}) + \sum_{j \in J}(1 - y_j)v_j + \sum_{j \in J} \lambda_j \left(y_j - \sum_{k=1}^{N} c_{jk}x_{jk}\right). \qquad (3)$$

It is a linear combination of the convex program's objective and constraints. Instead of prohibiting infeasible solutions (as done by the convex program), it charges a penalty for violated constraints (assuming positive $\lambda_j$). Now, the *dual function* of (CP) is defined as

$$g(\lambda) := \inf_{\substack{0 \preceq x \\ 0 \preceq y \preceq 1}} L(x, y, \lambda). \qquad (4)$$

An important property of the dual function $g$ is that for any $\lambda \succeq 0$, the value $g(\lambda)$ is a lower bound on the optimal value of (CP). Moreover, since (CP) is a relaxation of (IMP), $g(\lambda)$ is also a lower bound on the optimal value of (IMP). See the book by Boyd and Vandenberghe [8] for further details on these and similar known facts about (convex) optimization problems.

## 2.2 Power Consumption in Atomic Intervals

Let us give a more detailed description of the function $\mathcal{P}_k(x_{1k}, x_{2k}, \ldots, x_{nk})$. We defined $\mathcal{P}_k$ implicitly by mapping a given work assignment $x_{1k}, x_{2k}, \ldots, x_{nk}$ for interval $T_k$ to the power consumption of Chen et al.'s algorithm [11] during $T_k$. This guarantees an energy-minimal

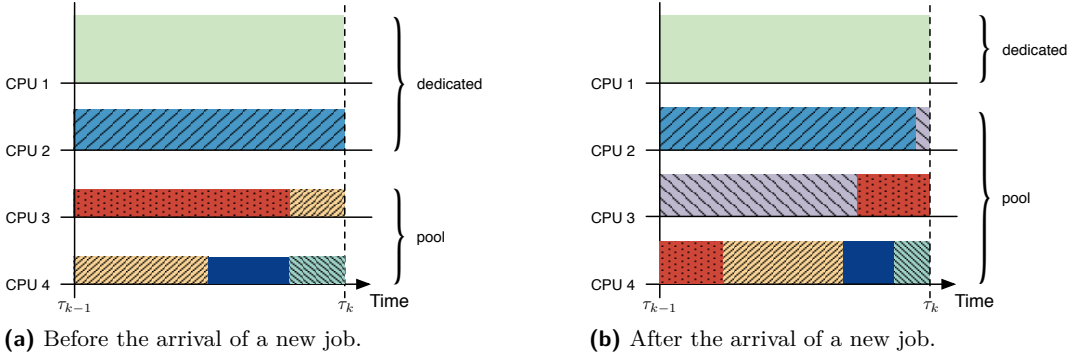

**(a)** Before the arrival of a new job.　**(b)** After the arrival of a new job.

**Figure 2** Schedules computed by Chen et al.'s algorithm before and after the arrival of a new job.

schedule for the given work assignment. In the following, we give a concise description of this algorithm and derive a more explicit formulation as well as some properties of $\mathcal{P}_k$.

To ease the discussion, let us assume that the jobs are numbered such that $x_{1k}w_1 \geq x_{2k}w_2 \geq \cdots \geq x_{nk}w_n$. In a nutshell, Chen et al.'s algorithm can be described as follows. Define the job set

$$\psi(k) := \left\{ j \in J \,\middle|\, j \leq m \,\wedge\, x_{jk} > 0 \,\wedge\, x_{jk}w_j \geq \frac{\sum_{j' > j} x_{j'k}w_{j'}}{m - j} \right\}. \tag{5}$$

These jobs are called *dedicated jobs* and are scheduled on their own *dedicated processor* using the energy-optimal (since minimal) speed $s_{jk} := \frac{x_{jk}w_j}{l_k}$. All remaining jobs, called *pool jobs*, are scheduled on the remaining *(pool) processors* in a greedy manner. The intuition is that dedicated jobs are larger than the remaining average workload and thus must be processed on a dedicated processor. See [7, Section 3.1] for a relatively short but more detailed description of the algorithm. Figure 2 illustrates the resulting schedule and how it may change due to the arrival of a new job. Using the above definition of dedicated jobs we can write $\mathcal{P}_k$ as

$$\mathcal{P}_k(x_{1k}, \ldots, x_{nk}) = \sum_{j \in \psi(k)} l_k \, \mathrm{P}_\alpha\!\left(\frac{x_{jk}w_j}{l_k}\right) + (m - |\psi(k)|) l_k \, \mathrm{P}_\alpha\!\left(\frac{\sum_{j \notin \psi(k)} x_{jk}w_j}{(m - |\psi(k)|) l_k}\right). \tag{6}$$

The following proposition gathers some important properties concerning the power consumption function $\mathcal{P}_k$ of an atomic interval $T_k$.

▶ **Proposition 1.** *Consider an arbitrary atomic interval $T_k$ together with its power consumption function $\mathcal{P}_k \colon \mathbb{R}^n_{\geq 0} \to \mathbb{R}$. This function has the following properties:*

*(a) It is convex and $\mathcal{P}_k(0) = 0$.*
*(b) It is differentiable with partial derivatives $\frac{\partial \mathcal{P}_k}{\partial x_{jk}}(x_{1k}, \ldots, x_{nk}) = w_j \cdot \mathrm{P}'_\alpha(s_{jk})$. Here, $s_{jk}$ denotes the speed used to schedule the workload $x_{jk}w_j$ in Chen et al.'s algorithm:*

$$s_{jk} = \begin{cases} x_{jk}w_j/l_k & , \text{ if } j \text{ is a dedicated job} \\ \frac{\sum_{j \notin \psi(k)} x_{jk}w_j}{(m - |\psi(k)|) l_k} & , \text{ if } j \text{ is a pool job.} \end{cases} \tag{7}$$

**Proof Sketch.** (a) The equality $\mathcal{P}_k(0) = 0$ is obvious from the definition of $\mathcal{P}_k$. The convexity follows easily from [7, Lemma 3.2]. There, the authors proved the convexity of $(x_{1k}, \ldots, x_{nk}) \mapsto \mathcal{P}_k(x_{1k}/w_1, \ldots, x_{nk}/w_n)$ (a linear transformation of $\mathcal{P}_k$).

(b) Differentiability is obvious for all points $(x_{1k}, \ldots, x_{nk})$ for which all the inequalities $x_{jk} w_j > \sum_{j' \geq j} x_{j'k} w_{j'}/(m-j)$ in Equation (2.2) are strict: For these, we have a small interval around $x_{jk}$ such that the set $\psi(k)$ of dedicated jobs does not change. On these intervals, $\mathcal{P}_k$ is essentially a linear map of the differentiable function $P_\alpha(s) = s^\alpha$. For other points, one can compute the left and right derivatives in $x_{jk}$, distinguishing whether job $j$ switched between a dedicated processor and a pool processor, whether $j$ stays on a dedicated processor, or whether $j$ stays on a pool processor and some other jobs switch between processor types. All cases yield the same left and right derivatives as given in the statement. ◂

We will also need to compare the result of Chen et al.'s algorithm before and after the arrival of a new job (cf. Figure 2). That is, how can the workloads on the processors change when a single entry of the work assignment changes from zero to some positive value?

▶ **Proposition 2.** *Consider Chen et al.'s algorithm called for some interval $T_k$ with the two work assignments $x = (x_1, x_2, \ldots, x_n, 0)$ and $x' = (x_1, x_2, \ldots, x_n, z)$ (i.e., before and after the arrival of a new job). Let $L_i$ and $L'_i$ denote the total workload on the $i$-th fastest processor in the resulting schedules, respectively. Then, we have $0 \leq L'_i - L_i \leq z$.*

**Proof Sketch.** We consider only the normalized case. That is, the case of unit workloads ($w_j = 1$ for all jobs) and an atomic interval of unit length ($l_k = 1$). The general case follows by a straightforward adaption. Without loss of generality, we furthermore assume $x_1 \geq x_2 \geq \cdots \geq x_n$. Note that we do not presume any relation between the newly arrived workload $z$ and the remaining workloads. Let $S$ and $S'$ be the schedules produced by Chen et al.'s algorithm for the work assignments $x$ and $x'$, respectively. Similarly, we use $d$ and $d'$ to denote the number of dedicated processors, and $L_{\text{pool}}$ and $L'_{\text{pool}}$ for the workload of a pool processor in $S$ and $S'$, respectively. Remember that pool processors have the smallest workload. That is, we have $L_i \geq L_{\text{pool}}$ and $L'_i \geq L'_{\text{pool}}$ for all $i \in \{1, 2, \ldots, m\}$.

We start with the proof of $L'_i - L_i \geq 0$. Observe that the arrival of the workload $z$ will not cause any of the former pool jobs to become a dedicated job (cf. Equation (5)). Moreover, by the same equation, for each dedicated processor that becomes a pool processor we also get a new pool job that has a workload of at least $L_{\text{pool}}$. Thus, the workload of pool processors from $S$ can only increase. The workload of the $i$-th fastest dedicated processor in $S$ is exactly $x_i$. If it becomes a pool processor, we have $x_i < L'_{\text{pool}} = L'_i$, yielding $L_i = x_i < L'_i$. If it stays a dedicated processor, its workload is the $i$-th largest value in $\{x_1, \ldots, x_n\}$ and, thus, at least as large as the $i$-th largest value in $\{x_1, \ldots, x_n\}$, yielding $L_i \leq L'_i$. To prove the second statement, $L'_i - L_i \leq z$, let us assume $L'_i - L_i > z$ and seek a contradiction. We distinguish two cases, depending on the type (pool or dedicated) of the $i$-th fastest processor in $S'$:

**processor $i$ is a pool processor in $S'$** Note that $z < L'_i - L_i \leq L'_i$ and $i$ being a pool processor implies that $z$ is also scheduled on a pool processor (cf. Equation (5)). As $d'$ is the number of dedicated processors, we must have $i > d'$. Moreover, all the jobs with workload less than $L'_{d'}$ must be pool jobs in $S'$. These are exactly the jobs which are scheduled on the processors $d' + 1, \ldots, m$ in schedule $S$. Thus, the total workload of all pool processors in $S'$ equals $(m - d')L'_i = z + \sum_{j > d'} L_j$. Using $i > d'$, $L'_{i'} - L_{i'} \geq 0$ for all $i' \in \{1, 2, \ldots, m\}$, and that all pool processors in $S'$ have the same workload, we get $z = (m - d')L'_i - \sum_{j > d'} L_j = \sum_{j > d'} (L'_i - L_j) = \sum_{j > d'} (L'_j - L_j) \geq L'_i - L_i$. This contradicts our assumption.

**processor $i$ is a dedicated processor in $S'$** Our assumption implies $L'_i > L_i + z \geq z$. Together with $i$ being a dedicated processor this yields $L'_i = x_i$ (because $x_i$ remains the $i$-th largest value in $\{x_1, x_2, \ldots, x_n, z\}$). But the assumption also implies $L'_i > L_i + z \geq L_i \geq x_i$. We get the contradiction $x_i = L'_i > x_i$. ◂

```
1   {executed each time a new job j ∈ J arrives}
2   init  x_{jk}, y_j, and λ_j with zero for all k ∈ { 1, 2, . . . , N }
3   compute λ_{jk} := δ ∂P_k/∂x_{jk} (x_{1k}, x_{2k}, . . . , x_{jk}, 0, . . . , 0) for each interval T_k ⊆ [r_j, d_j)
4
5   let the set 𝒯_min contain all T_k with minimal λ_{jk}
6   for each T_k ∈ 𝒯_min in parallel:
7       increase x_{jk} in a continuous way (which in turn raises λ_{jk} according to line 3)
8       ensure that all λ_{jk} of intervals in 𝒯_min remain equal
9       update 𝒯_min whenever the λ_{jk} reach a λ_{jk'} with T_{k'} ∉ 𝒯_min
10      stop increasing once one of the following comes true
11      (a)  ∑ x_{jk} = 1 :   set    y_j := 1, λ_j := λ_{jk}
12      (b)  λ_{jk} = v_j :   reset  x_{jk} := 0, λ_j := λ_{jk}
```

▮ **Listing 1** Primal-Dual Algorithm PS with parameter $\delta$.

## 3 An Online Greedy Primal-Dual Algorithm

The goal of this section is to use the convex programming formulation (CP) and its dual function $g\colon \mathbb{R}^n \to \mathbb{R}$ to derive a provably good online algorithm for our scheduling problem. We start by describing an algorithm that computes a solution to (CP) in an online fashion, but knowing the time partitioning $T_k$ ($k \in \{1, 2, \ldots, n\}$). Subsequently, we explain how this solution is used to compute the actual schedule and how we handle the fact that the actual atomic intervals are not known beforehand. To solve (CP), we use a greedy primal-dual approach for convex programs as suggested by Gupta, Krishnaswamy, and Pruhs [12]. Our algorithm extends their framework to the multiprocessor case and to profitable scheduling models. It shows how to incorporate rejection policies into the framework (handling the integral constraints in the convex program) and how to cope with more complex power functions of a system (in our case $\mathcal{P}_k$).

**The Primal-Dual Algorithm.**

Our primal-dual algorithm, in the following referred to as PD, maintains a set of primal variables $(x, y)$ and a set of dual variables $\lambda$, all initialized with zero. Whenever a new job (i.e., a constraint in (CP)) arrives, we start to increase the primal variables $x_{jk}$ ($k \in \{1, 2, \ldots, N\}$) in a greedy fashion until either the full job is scheduled (i.e., $\sum_k x_{jk} = 1$) or the planned energy investment for job $j$ becomes too large compared to its value. In the latter case, the variables $x_{jk}$ are reset to zero, $\lambda_j$ is set to $v_j$, and $y_j$ remains zero (the job is rejected). Otherwise, we set $y_j$ to one (the job is finished) and $\lambda_j$ to essentially the current rate of cost increase per job workload. When greedily increasing the primal variables, we assign the next infinitesimal small portion of job $j$ to those atomic intervals that cause the smallest increase in costs. Essentially, these are the intervals where $j$'s workload would be scheduled with the slowest speed. See Listing 1 for the algorithm.

The described algorithm is similar to primal-dual algorithms known from linear programming, where primal and/or dual variables are raised at certain rates until the (relaxed) complementary slackness conditions are met. In fact, this algorithm is derived by using relaxed versions of the Karush-Kuhn-Tucker (KKT) conditions, essentially a generalization of the complementary slackness conditions for convex (or even general nonlinear) programs. The actual schedule used is the one computed by Chen et al.'s algorithm when applied to the current work assignment given by the primal variables $x_{jk}$ for the atomic interval $T_k$.

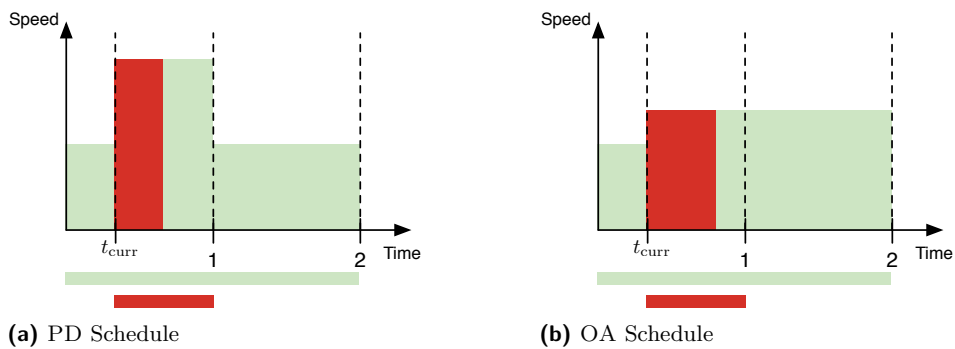

**(a)** PD Schedule

**(b)** OA Schedule

**Figure 3** The dashed lines indicate atomic intervals; the bars below, the jobs' availability. Note that PD's schedule is more conservative in comparison, leaving more room for scheduling jobs that might occur during the last atomic interval.

### Concerning the Time Partitioning.

Our algorithm formulation assumes a priori knowledge of the atomic intervals $T_k$. However, since the jobs arrive in an online fashion, the exact partitioning is actually not known to the algorithm. One can reformulate the algorithm such that it uses the intervals $T'_k$ induced by the jobs $J' = \{1, 2, \ldots, j\} \subseteq J$ it knows so far. If a refinement of an atomic interval $T'_k = T_{k_1} \cup T_{k_2}$ occurs due to the arrival of a new job, the already assigned job portions are simply split according to the ratios $|T_{k_1}|/|T'_k|$ and $|T_{k_2}|/|T'_k|$. This reformulated algorithm produces an identical schedule. To see this, note that the algorithm with a priori knowledge of the refinement $T'_k = T_{k_1} \cup T_{k_2}$ treats both intervals $T_{k_1}$ and $T_{k_2}$ as identical (with respect to their relative size $|T_{k_i}|/|T'_k|$) up to the point when the job causing the refinement arrives.

### Relation to the OA Algorithm.

For the case of a single processor and sufficiently high job values, algorithm PD is quite similar to the popular OA algorithm by Yao, Demers, and Shenker [14]. When a new job arrives, PD essentially finds the atomic intervals of lowest speed and increases their speed to free computational resources to be used for the new job. This is also true for the OA algorithm. However, while PD never changes how other jobs are distributed over atomic intervals, OA may actually influence this distribution. Figure 3 gives a simple example for the structural difference of the resulting schedules. Another interesting observation is that, in the single processor case, our analysis yields the very same optimal rejection policy as an OA-based algorithm by Chan, Lam, and Li [10]. Indeed, as we will see in Section 4, our analysis yields that $\delta = \alpha^{1-\alpha}$ is the optimal choice for the parameter $\delta$. Using this parameter, one can easily check that our rejection policy essentially states to reject a job if its energy consumption in the planned schedule exceeds $\alpha^{\alpha-2} \cdot v_j$. Or, equivalently, a job is rejected if its speed in the planned schedule exceeds $\alpha^{\frac{\alpha-2}{\alpha-1}} \cdot (v/w)^{\alpha-1}$, the rejection policy from [10].

## 4 Analysis

In the following, let $(\tilde{x}, \tilde{y})$ and $\tilde{\lambda}$ denote the primal and dual variables computed by our algorithm PD. Remember that the final schedule computed by PD is derived by applying Chen et al.'s algorithm to the $\tilde{x}_{1k}, \ldots, \tilde{x}_{nk}$ values in each atomic interval $T_k$. We refer to this schedule as the $(\tilde{x}, \tilde{y})$-schedule or simply as the schedule PD. Our goal is to use $g(\tilde{\lambda})$ to bound the cost of this schedule (referred to as cost(PD)). Our main result is

▶ **Theorem 3.** *The competitive ratio of algorithm* PD *with the parameter $\delta$ set to $\frac{1}{\alpha^{\alpha-1}}$ is at most $\alpha^\alpha$. Moreover, there is a problem instance for which* PD *is exactly by a factor of $\alpha^\alpha$ worse than an optimal algorithm. That is, our upper bound is optimal.*

For the upper bound, we show that $\text{cost}(\text{PD}) \leq \alpha^\alpha g(\tilde{\lambda})$. Since, by duality, $g(\tilde{\lambda})$ is also a lower bound on the optimal value of (CP) and, thereby, on the optimal value of (IMP), we get $\frac{\text{cost}(\text{PD})}{\text{cost}(\text{OPT})} \leq \alpha^\alpha$. The lower bound follows from a known result for traditional energy-efficient scheduling (without job values but the necessity to finish all jobs) by setting the job values sufficiently high.

In the remainder, we develop the key ingredients for the proof of Theorem 3. We start in Section 4.1 and derive a more explicit formulation of the dual function value $g(\tilde{\lambda})$ by relating it to a certain (infeasible) solution to our convex program (CP) and a corresponding schedule. Section 4.2 further simplifies this formulation by expressing $g(\tilde{\lambda})$ solely in terms of the jobs (instead of their workloads in different atomic intervals). Based on this job-centric formulation, Section 4.3 develops different bounds for the dual function value depending on certain job characteristics. The actual proof of Theorem 3 combines these bounds and can be found in Section 4.4.

## 4.1 Structure of an Optimal Infeasible Solution

First of all, note that the value $g(\tilde{\lambda}) = \inf L(x, y, \tilde{\lambda})$ (cf. Equation (4)) is finite and obtained by a pair $(\hat{x}, \hat{y})$ of primal variables. These primal variables can be interpreted as a (possibly infeasible) solution to the convex program (CP). Moreover, for our fixed dual variable $\tilde{\lambda}$, this solution is optimal in that it minimizes the sum of the objective cost and the penalty for violated constraints. In this sense, we refer to $(\hat{x}, \hat{y})$ as an *optimal infeasible solution*. Our goal is to understand the structure of this solution, which will eventually allow us to write $g(\tilde{\lambda})$ in a more explicit way. The results of this subsection are related to results from [12], but more involved due to the more complex nature of our objective function.

Note that $\hat{x}$ and $\hat{y}$ may differ largely from $\tilde{x}$ and $\tilde{y}$. However, the following lemmas show a strong correlation between this optimal infeasible solution and the feasible (partially integral) solution computed by algorithm PD.

▶ **Lemma 4.** *Consider an optimal infeasible solution $(\hat{x}, \hat{y})$. Without loss of generality, we can assume that it has the following properties:*

*(a) $\hat{y} = \tilde{y}$*
*(b) For any atomic interval $T_k$, there are at most $m$ different jobs $j$ with $\hat{x}_{jk} > 0$.*

**Proof.** (a) Consider an arbitrary job $j \in J$ and remember that the domain for the variables $\hat{y}_j$ is restricted to $[0, 1]$. The contribution of variable $\hat{y}_j$ to $g(\tilde{\lambda}) = L(\hat{x}, \hat{y}, \tilde{\lambda})$ is exactly $\hat{y}_j(\tilde{\lambda}_j - v_j)$, as can be seen by considering Equation (3). If $\tilde{\lambda}_j < v_j$, this is minimized by choosing $\hat{y}_j$ maximal ($\hat{y}_j = 1$). Otherwise, we must have $\tilde{\lambda}_j = v_j$ (by the definition of algorithm PD). This allows us to choose $\hat{y}_j$ arbitrarily, such that we can set it to zero. Both choices correspond exactly to the way $\tilde{y}_j$ is set by algorithm PD.
(b) Assume there are more than $m$ jobs with $\hat{x}_{jk} > 0$. We can assume $c_{jk} = 1$ for these jobs, because otherwise we could set $\hat{x}_{jk} = 0$ without increasing $g(\tilde{\lambda}) = L(\hat{x}, \hat{y}, \tilde{\lambda})$. Now, the values $\hat{x}_{1k}, \ldots, \hat{x}_{nk}$ correspond to a work assignment for the atomic interval $T_k$, as used by Chen et al.'s algorithm (cf. Section 2.2). By Equation (3), the contribution of these values to $g(\tilde{\lambda}) = L(\hat{x}, \hat{y}, \tilde{\lambda})$ is given by $\mathcal{P}_k(\hat{x}_{1k}, \ldots, \hat{x}_{nk}) - \sum_{j \in J} \tilde{\lambda}_j \hat{x}_{jk}$. Since there are more than $m$ jobs $j$ with nonzero $\hat{x}_{jk}$, at least two of them must share a processor in the schedule computed by Chen et al.'s algorithm for this work assignment. In other words, there are two pool jobs $j, j' \in J \setminus \psi(k)$ with $\hat{x}_{jk}, \hat{x}_{j'k} > 0$. Together with Equation (6), we see that the contribution of $\hat{x}_{jk}$ and $\hat{x}_{j'k}$ to $g(\tilde{\lambda})$ consists of two terms: a convex term

$$(m - |\psi(k)|)l_k \operatorname{P}_\alpha\left(\frac{\sum_{j \notin \psi(k)} \hat{x}_{jk} w_j}{(m - |\psi(k)|)l_k}\right)$$

and a linear term $-\tilde{\lambda}_j \hat{x}_{jk} - \tilde{\lambda}_{j'} \hat{x}_{j'k}$. By changing $\hat{x}_{jk}$ and $\hat{x}_{j'k}$ along the line that keeps the sum $\hat{x}_{jk} w_j + \hat{x}_{j'k} w_{j'}$ constant, we can decrease one of the variables (say $\hat{x}_{jk}$) and increase the other such that the first (convex) term remains constant and the second (linear) term is not increased. This will not effect the type (dedicated or pool) of other jobs. The only job that may change its type is job $j'$, as it may become a dedicated job. Once this happens, we iterate the process with two other pool jobs. As the number of dedicated jobs is upper bounded by $m$, this can happen only finitely often. Thus, at some point we can decrease $\hat{x}_{jk}$ all the way to zero without increasing the dual function value $g(\tilde{\lambda})$. We continue eliminating $\hat{x}_{jk}$ variables until at most $m$ of them are nonzero. ◂

Given an atomic interval $T_k$, we call the jobs $j$ with $\hat{x}_{jk} > 0$ the *contributing jobs of $T_k$* and denote the corresponding job set by $\varphi(k)$. As done in the proof of Lemma 4, we can consider $\hat{x}$ as a work assignment for the atomic intervals $T_k$. By applying Chen et al.'s algorithm, we get a schedule whose energy cost in interval $T_k$ is exactly $\mathcal{P}_k(\hat{x}_{1k}, \ldots, \hat{x}_{nk})$. We refer to this schedule as the $(\hat{x}, \hat{y})$-schedule. Using this terminology, the second statement of Lemma 4 essentially says that in this schedule at most $m$ jobs are scheduled in any atomic interval $T_k$. Moreover, it follows immediately from the description of Chen et al.'s algorithm that all contributing jobs are dedicated jobs of the corresponding atomic interval.

We can derive a slightly more explicit characterization of the contributing jobs $\varphi(k)$ of an atomic interval $T_k$ by exploiting that $(\hat{x}, \hat{y})$ is a minimizer of $(x, y) \mapsto L(x, y, \tilde{\lambda})$.

▶ **Lemma 5.** *Consider any atomic interval $T_k$ and its contributing jobs $\varphi(k)$. Define the value $\hat{s}_j := (\tilde{\lambda}_j / \alpha w_j)^{\frac{1}{\alpha - 1}}$ for any job $j$.*

(a) *For any $j \in \varphi(k)$ we have $\hat{x}_{jk} = \frac{l_k}{w_j} \hat{s}_j = \frac{l_k}{w_j} (\tilde{\lambda}_j / \alpha w_j)^{\frac{1}{\alpha - 1}}$. Moreover, $j$ is scheduled at constant speed $\hat{s}_j$ in the $(\hat{x}, \hat{y})$-schedule.*

(b) *The total contribution of the $\hat{x}_{jk}$ variables to the dual function value $g(\tilde{\lambda})$ is*

$$(1 - \alpha)l_k \sum_{j \in \varphi(k)} \left(\frac{\hat{x}_{jk} w_j}{l_k}\right)^\alpha = (1 - \alpha)l_k \sum_{j \in \varphi(k)} \hat{s}_j^\alpha. \tag{8}$$

(c) *Let $n_k$ denote the number of jobs available in the atomic interval $T_k$ (i.e., jobs with $c_{jk} = 1$). The contributing jobs $\varphi(k)$ are the $\min(m, n_k)$ jobs with maximal $\hat{s}_j$-values under all available jobs.*

**Proof.** (a) By definition, $\hat{x}$ is a minimizer of $x \mapsto L(x, \hat{y}, \tilde{\lambda})$. This implies that we must have $\frac{\partial L}{\partial x_{jk}}(\tilde{\lambda}, \hat{x}, \hat{y}) = 0$ for any contributing job $j \in \varphi(k)$. We get

$$0 = \tfrac{\partial L}{\partial x_{jk}}(\tilde{\lambda}, \hat{x}, \hat{y}) = \tfrac{\partial \mathcal{P}_k}{\partial x_{jk}}(\hat{x}_{1k}, \ldots, \hat{x}_{nk}) - \tilde{\lambda}_j$$

$$= w_j \cdot \operatorname{P}'_\alpha\left(\frac{\hat{x}_{jk} w_j}{l_k}\right) - \tilde{\lambda}_j = \alpha w_j \left(\frac{\hat{x}_{jk} w_j}{l_k}\right)^{\alpha - 1} - \tilde{\lambda}_j,$$

which yields the first statement by rearranging. The second statement follows from this by noticing that $\frac{\hat{x}_{jk}w_j}{l_k}$ is the speed used by Chen et al.'s algorithm for the (dedicated) job $j$.

(b) By definition of $g(\tilde{\lambda}) = L(\hat{x}, \hat{y}, \tilde{\lambda})$, we get that the total contribution of the $\hat{x}_{jk}$ variables is (there are no pool jobs!)

$$\mathcal{P}_k(\hat{x}_{1k}, \ldots, \hat{x}_{nk}) - \sum_{j \in \varphi(k)} \tilde{\lambda}_j \hat{x}_{jk} = \sum_{j \in \varphi(k)} l_k \, \mathrm{P}_\alpha\!\left(\frac{\hat{x}_{jk}w_j}{l_k}\right) - \sum_{j \in \varphi(k)} \tilde{\lambda}_j \hat{x}_{jk}$$

$$= l_k \sum_{j \in \varphi(k)} \mathrm{P}_\alpha(\hat{s}_j) - \alpha l_k \sum_{j \in \varphi(k)} \frac{\tilde{\lambda}_j}{\alpha w_j} \hat{s}_j = (1-\alpha) l_k \sum_{j \in \varphi(k)} \hat{s}_j^\alpha.$$

(c) The contributing jobs must be chosen such that their contribution is minimized. Using statement (b) and $\alpha > 1$, we see that this is the case when choosing the maximal number of available jobs (at most $m$) with the largest $\hat{s}_j$-values.  ◀

## 4.2 A Job-centric Formulation of the Dual Function

In the following, we assume that the optimal infeasible solution $(\hat{x}, \hat{y})$ adheres to Lemma 4. That is, we have $\hat{y} = \tilde{y}$ and we can relate the optimal infeasible solution to the $(\hat{x}, \hat{y})$-schedule which schedules in each atomic interval $T_k$ exactly the $|\varphi(k)|\ (\leq m)$ available jobs with the largest $\hat{s}_j = (\tilde{\lambda}_j/\alpha w_j)^{\frac{1}{\alpha-1}}$-values, each on its own dedicated processor at speed $\hat{s}_j$. We use the somewhat lax notation $k \in \varphi^{-1}(j)$ to refer to the atomic intervals $T_k$ to which $j$ contributes. Our main goal in this section is to derive a formulation of the dual function value solely in terms of the jobs. We will also define and discuss the *trace* of a job, which helps to relate any job (even if unfinished) to a certain amount of energy consumed by our PD algorithm.

Given a job $j \in J$, let $l(j) := \sum_{k \in \varphi^{-1}(j)} l_k$ denote the total time it is scheduled in the $(\hat{x}, \hat{y})$-schedule. Moreover, let $E_{\tilde{\lambda}}(j)$ denote the total energy invested by the $(\hat{x}, \hat{y})$-schedule into job $j$. Now, we can formulate the following lemma.

▶ **Lemma 6.** *For any job $j \in J$, the total energy invested by the optimal infeasible solution into job $j$ is $E_{\tilde{\lambda}}(j) = l(j)\hat{s}_j^\alpha$. Moreover, the dual function value $g(\tilde{\lambda})$ can be written as*

$$g(\tilde{\lambda}) = (1-\alpha) \sum_{j \in J} E_{\tilde{\lambda}}(j) + \sum_{j \in J} \tilde{\lambda}_j. \tag{9}$$

**Proof.** The equality $E_{\tilde{\lambda}}(j) = l(j)\hat{s}_j^\alpha$ follows immediately from the above definitions, as $j$ is processed by the $(\hat{x}, \hat{y})$-schedule at constant speed $\hat{s}_j$ for a total time of exactly $l(j)$. For the lemma's main statement, remember that $\hat{y}_j = 0$ if and only if $\tilde{\lambda}_j = v_j$. Otherwise we have $\hat{y}_j = 1$. Thus, the contribution of $\hat{y}_j$ to $g(\tilde{\lambda})$ is exactly $(1-\hat{y}_j)v_j + \tilde{\lambda}_j \hat{y}_j = \tilde{\lambda}_j$. As we have seen in Lemma 5 for a fixed $k$, the contribution of all $\hat{x}_{jk}$ to $g(\tilde{\lambda})$ is exactly $(1-\alpha)l_k \sum_{j \in \varphi(k)} \hat{s}_j^\alpha$. Summing over all $k$, we get that the total contribution of the $\hat{x}$-variables equals

$$\sum_{k=1}^N (1-\alpha) l_k \sum_{j \in \varphi(k)} \hat{s}_j^\alpha = (1-\alpha) \sum_{k=1}^N \sum_{j \in \varphi(k)} l_k \hat{s}_j^\alpha = (1-\alpha) \sum_{j \in J} \sum_{k \in \varphi^{-1}(j)} l_k \hat{s}_j^\alpha$$

$$= (1-\alpha) \sum_{j \in J} l(j)\hat{s}_j^\alpha = (1-\alpha) \sum_{j \in J} E_{\tilde{\lambda}}(j).$$

Summing up the contributions of the $\hat{x}$- and $\hat{y}$-variables we get the desired statement.  ◀

**Tracing a Job.**

Given a job $j$, we define its *trace* as a set of tuples $(T_k, i)$ with $k \in \{1, 2, \ldots, N\}$ and $i \in \{1, 2, \ldots, m\}$. That is, a set of atomic intervals, each coupled with a certain processor. Our goal is to choose these such that we can account the energy $E_{\tilde{\lambda}}(j)$ used in the optimal infeasible solution on job $j$ to the energy used by algorithm PD during $j$'s trace (on the coupled processors). For the formal definition, let us first partition the contributing jobs $\varphi(k)$ of an interval $T_k$ into the subset $\varphi_1(k) := \{j \in \varphi(k) \mid \tilde{y}_j = 1\}$ of jobs finished by PD and the subset $\varphi_2(k) := \{j \in \varphi(k) \mid \tilde{y}_j = 0\}$ of jobs unfinished by PD. Now, for any job $j \in J$ we define its *trace* $\mathrm{Tr}(j)$ as follows:

**Case $\tilde{y}_j = 1$:** $(T_k, i) \in \mathrm{Tr}(j) \iff \hat{s}_j$ is the $i$-th largest value[1] in $\{\hat{s}_{j'} \mid j' \in \varphi_1(k)\}$
**Case $\tilde{y}_j = 0$:** $(T_k, |\varphi_1(k)| + i) \in \mathrm{Tr}(j) \iff \hat{s}_j$ is the $i$-th largest value in $\{\hat{s}_{j'} \mid j' \in \varphi_2(k)\}$

That is, jobs that are finished by PD are mapped to the fastest processors in each atomic interval $T_k$ for which they are contributing jobs, in decreasing order of their $\hat{s}_j$-values. Jobs contributing to $T_k$ but which are unfinished by PD are mapped to the remaining processors (the exact order is not important in this case). Note that by this mapping, all traces $\mathrm{Tr}(j)$ are pairwise disjoint. We use the notation $E_{\mathrm{PD}}(j)$ to refer to the power consumption of PD during $j$'s trace. That is, the power consumption on the $i$-th fastest processor in the atomic interval $T_k$ for any $(T_k, i) \in \mathrm{Tr}(j)$. We use $E_{\mathrm{PD}}$ to denote the total power consumption of PD. Since the job traces are pairwise disjoint, we obviously have $E_{\mathrm{PD}} \geq \sum_{j \in J} E_{\mathrm{PD}}(j)$.

The following proposition formulates an important structural property of a job's trace. It gives us different lower bounds on the speed used by PD during a job's trace, depending on whether it is finished or not. To this end, let $\tilde{s}_j$ denote the speed PD planned to use for job $j$ just before $\tilde{\lambda}_j$ got fixed (i.e., just before PD decides whether to finish $j$ or not). If $j$ is finished, we have (cf. algorithm description and Proposition 1)

$$\tilde{\lambda}_j = \delta \frac{\partial \mathcal{P}_k}{\partial x_{jk}}(\tilde{x}_{1k}, \ldots, \tilde{x}_{jk}, 0, \ldots, 0) = \delta w_j \, \mathrm{P}'_\alpha(\tilde{s}_j). \tag{10}$$

Solving this for $\tilde{s}_j$ yields $\tilde{s}_j = (\tilde{\lambda}_j / \delta \alpha w_j)^{1/\alpha - 1} = \delta^{-1/\alpha - 1} \hat{s}_j$. Similarly, we also get $\tilde{s}_j = \delta^{-1/\alpha - 1} \hat{s}_j$ for unfinished jobs. We use $\check{x}_j = \sum \check{x}_{jk} < 1$ to denote the corresponding portions of the unfinished job $j$ planned to be scheduled by PD just before $j$ was rejected.

▶ **Proposition 7.** *Consider $(T_k, i) \in \mathrm{Tr}(j)$ for a job $j \in J$. Let $s(i, k)$ denote the speed of the $i$-th fastest processor during $T_k$ in the final schedule computed by* PD. *Then:*

*(a) If $j$ is finished by* PD, *then $s(i, k) \geq \tilde{s}_j$.*
*(b) If $j$ is not finished by* PD, *then $s(i, k) \geq \tilde{s}_j - \frac{\check{x}_{jk} w_j}{l_k}$.*

**Proof.** (a) Remember that $\tilde{s}_j = \delta^{-1/\alpha - 1} \hat{s}_j$. Because of this relation and the definition of $(T_k, i) \in \mathrm{Tr}(j)$, we must have that $\tilde{s}_j$ is the $i$-th largest value in $\{\tilde{s}_{j'} \mid j' \in \varphi_1(k)\}$. Together with Lemma 5(c), we even have that $\tilde{s}_j$ is the $i$-th largest value under all available jobs finished by PD. At the time $\tau_{k-1}$ (the start of interval $T_k$), all these available jobs $j'$ have arrived. We consider two cases: If $j$ is a dedicated job at this time, it is scheduled with a speed of exactly $\tilde{s}_j$. Moreover, all the $i - 1$ available jobs $j'$ with $\tilde{s}_{j'} \geq \tilde{s}_j$ are dedicated jobs and are scheduled with a speed of $\tilde{s}_{j'}$, respectively. Thus, $j$ is scheduled on the $i$-th fastest processor, yielding $s(i, k) \geq \tilde{s}_j$. If $j$ is a pool job at this time, it is scheduled on one of the pool processors at a speed of at least $\tilde{s}_j$. But then, since pool processors are the slowest processors, the $i$-th fastest processor must also run at a speed of at least $\tilde{s}_j$.

---

[1] Ties are resolved arbitrarily but consistently.

(b) Remember that $\check{x}_{jk}$ denotes the portion of job $j$ PD planned to schedule in $T_k$ just before $j$ got rejected. If $j$ was planned as a dedicated job, we have $l_k \tilde{s}_j = \check{x}_{jk} w_j$. This trivially yields the desired statement because of $s(i,k) \geq 0$. If $j$ was not planned as a dedicated job, it was to be processed on a pool processor. Let $L(i,k)$ denote the workload on the $i$-th fastest processor during $T_k$ just after $j$ was rejected (i.e., without $\check{x}_{jk} w_j$). Similarly, let $L'(i,k)$ denote the workload on the $i$-th fastest processor during $T_k$ just before $j$ was rejected (i.e., including $\check{x}_{jk} w_j$). Proposition 2 gives us $L'(i,k) - L(i,k) \leq \check{x}_{jk} w_j$. Moreover, since $j$ was planned as a pool job (which run at minimal speed), we must have $l_k \tilde{s}_j \leq L'(i,k)$. Combining these inequalities yields that the speed $L(i,k)/l_k$ on the $i$-th fastest processor during $T_k$ at $j$'s arrival was at least $\tilde{s}_j - \frac{\check{x}_{jk} w_j}{l_k}$. As Proposition 2 also implies that the workload (and, thus, the speed) of the $i$-th fastest processor in an atomic interval can only increase due to the arrival of new jobs, we get the desired statement. ◀

## 4.3 Balancing the Different Cost Components

As our goal is to lower-bound the dual function value $g(\tilde{\lambda}) = (1-\alpha) \sum E_{\tilde{\lambda}}(j) + \sum \tilde{\lambda}_j$ by the cost of algorithm PD, we have to relate the values $E_{\tilde{\lambda}}(j)$ and $\tilde{\lambda}_j$ to the energy- and value- costs of PD. It depends on the job itself how this is done exactly. For example, in the case of finished jobs, both terms can be related to the actual energy consumption of PD in a relatively straightforward way. This becomes much harder if the job is not finished by PD: after all, in this case PD does not invest any energy into the job. The job's trace plays a crucial role in this case, as it allows us to account the energy investment of the optimal infeasible solution to the energy PD consumed during the trace. The next proposition gathers the most important relations to be used in the following proofs.

▶ **Proposition 8.** *Consider an arbitrary job $j \in J$:*

(a) $E_{\tilde{\lambda}}(j) = \tilde{\lambda}_j \frac{\hat{x}_j}{\alpha}$
(b) *If $j$ is finished by* PD, *then* $E_{\tilde{\lambda}}(j) \leq \delta^{\frac{\alpha}{\alpha-1}} E_{\mathrm{PD}}(j)$.
(c) *If $j$ is not finished by* PD *and $\hat{x}_j > \delta^{\frac{1}{\alpha-1}}$, then*

$$E_{\tilde{\lambda}}(j) < \delta^{\frac{\alpha}{\alpha-1}} \left(1 - \frac{\delta^{\frac{1}{\alpha-1}}}{\hat{x}_j}\right)^{-\alpha} E_{\mathrm{PD}}(j). \tag{11}$$

**Proof.** (a) We use the identities $\hat{s}_j = (\tilde{\lambda}_j/\alpha w_j)^{\frac{1}{\alpha-1}}$ and $l(j)\hat{s}_j = \hat{x}_j w_j$ (cf. Lemma 5) and compute

$$E_{\tilde{\lambda}}(j) = l(j)\hat{s}_j^\alpha = l(j)\hat{s}_j \cdot \hat{s}_j^{\alpha-1} = \hat{x}_j w_j \cdot \frac{\tilde{\lambda}_j}{\alpha w_j} = \tilde{\lambda}_j \frac{\hat{x}_j}{\alpha}.$$

(b) Assume $j$ is finished by PD. Remember that $\tilde{s}_j$ denotes the speed assigned to $j$ when it arrived and $\tilde{\lambda}_j$ got fixed. We have the relation $\tilde{s}_j = \delta^{-1/\alpha - 1} \hat{s}_j$ (cf. Section 4.2). Let $s_{\min}$ denote the minimal speed of $j$'s trace in the final $(\tilde{x}, \tilde{y})$-schedule produced by PD. That is, there is a tuple $(T_k, i) \in \mathrm{Tr}(j)$ such that the $i$-th fastest processor in $T_k$ runs at speed $s_{\min}$ and $E_{\mathrm{PD}}(j) \geq l(j) s_{\min}^\alpha$. By Proposition 7 we must have $s_{\min} \geq \tilde{s}_j$. We compute

$$E_{\tilde{\lambda}}(j) = l(j)\hat{s}_j^\alpha = \delta^{\frac{\alpha}{\alpha-1}} l(j)\tilde{s}_j^\alpha \leq \delta^{\frac{\alpha}{\alpha-1}} l(j) s_{\min}^\alpha \leq \delta^{\frac{\alpha}{\alpha-1}} E_{\mathrm{PD}}(j).$$

(c) Applying Proposition 7 to all $(T_k, i) \in \mathrm{Tr}(j)$ yields that the total workload $L$ that is processed by PD during $j$'s trace is at least $l(j)\tilde{s}_j - \check{x}_j w_j > l(j)\tilde{s}_j - w_j$. The minimum energy necessary to process this workload in $l(j)$ time units is $l(j)\left(L/l(j)\right)^\alpha$. We compute

$$E_{\text{PD}}(j) \geq l(j)\left(\frac{L}{l(j)}\right)^\alpha > l(j)\left(\frac{l(j)\tilde{s}_j - w_j}{l(j)}\right)^\alpha = l(j)\tilde{s}_j^\alpha \left(1 - \frac{w_j}{\tilde{s}_j l(j)}\right)^\alpha$$

$$= \delta^{-\frac{\alpha}{\alpha-1}} E_{\tilde{\lambda}}(j) \left(1 - \frac{\delta^{\frac{1}{\alpha-1}}}{\hat{x}_j}\right)^\alpha.$$

Rearranging the inequality yields the desired statement. ◀

Note that the bound for unfinished jobs in Proposition 8 has an additional factor $> 1$ compared to the one for finished jobs. However, for large enough $\hat{x}_j$ this factor becomes nearly one. Thus, we will apply this bound only in cases of large $\hat{x}_j$. If $\hat{x}_j$ is relatively small, we will instead bound $E_{\tilde{\lambda}}(j)$ only by its value. We continue by describing the different types of jobs we consider. In total, we differentiate between three job categories:

**Finished Jobs** These are all jobs $j$ with $\tilde{y}_j = 1$ (i.e., jobs finished by PD). As mentioned above, we bound both components $E_{\tilde{\lambda}}(j)$ and $\tilde{\lambda}_j$ of $g(\tilde{\lambda})$ by the actual energy consumption of PD. We use $J_1 \coloneqq \{\, j \in J \mid \tilde{y}_j = 1 \,\}$ to refer to this job category.

**Unfinished, Low-yield Jobs** We use the term *low-yield jobs* to refer to jobs not finished by PD and which have a relatively small $\hat{x}_j$. That is, jobs of which the optimal infeasible solution does not schedule too large a portion. Intuitively, the value of such jobs must be small, because otherwise it would have been beneficial to schedule a larger portion of them in the optimal infeasible solution. In this sense, these jobs are low-yield and we will exploit this fact by bounding both components $E_{\tilde{\lambda}}(j)$ and $\tilde{\lambda}_j$ of $g(\tilde{\lambda})$ by the job value PD is charged for not finishing $j$. More formally, this job category is defined as $J_2 \coloneqq \{\, j \in J \mid \tilde{y}_j = 0 \wedge \hat{x}_j \leq \frac{\alpha - \alpha^{1-\alpha}}{\alpha - 1} \,\}$.

**Unfinished, High-yield Jobs** Correspondingly, the term *high-yield jobs* refers to jobs finished by PD and which have a relatively large $\hat{x}_j$. More exactly, these jobs are given by $J_3 \coloneqq \{\, j \in J \mid \tilde{y}_j = 0 \wedge \hat{x}_j > \frac{\alpha - \alpha^{1-\alpha}}{\alpha - 1} \,\}$. This proves to be the most challenging case, as neither do the jobs feature a particularly small value nor does PD invest any energy into their execution. Instead, we use a mix of the job's value and the energy spent by PD during $j$'s trace to account for its contribution. One has to carefully balance what portions of $E_{\tilde{\lambda}}(j)$ and $\tilde{\lambda}_j$ to bound by either $E_{\text{PD}}(j)$ or by $v_j$.

In accordance with these job categories, we split the value of the dual function by the corresponding contributions. That is, $g(\tilde{\lambda}) = \sum_{i=1}^3 g_i(\tilde{\lambda})$, where $g_i(\tilde{\lambda}) = (1-\alpha)\sum_{j \in J_i} E_{\tilde{\lambda}}(j) + \sum_{j \in J_i} \tilde{\lambda}_j$. The following lemmas bound each contribution separately.

▶ **Lemma 9** (Finished Jobs). $g_1(\tilde{\lambda}) \geq \delta E_{\text{PD}} + (1-\alpha)\delta^{\frac{\alpha}{\alpha-1}} \sum_{j \in J_1} E_{\text{PD}}(j)$.

**Proof.** We have $g_1(\tilde{\lambda}) = (1-\alpha)\sum_{j \in J_1} E_{\tilde{\lambda}}(j) + \sum_{j_1 \in J} \tilde{\lambda}_j$. Using Proposition 8(b) and $\alpha > 1$ we bound the first summand by $(1-\alpha)\delta^{\frac{\alpha}{\alpha-1}} \sum_{j \in J_1} E_{\text{PD}}(j)$. For the second summand, we get

$$\sum_{j \in J_1} \tilde{\lambda}_j = \sum_{j \in J_1} \sum_{k=1}^N \tilde{x}_{jk}\tilde{\lambda}_j = \sum_{j \in J_1} \sum_{k=1}^N \tilde{x}_{jk}\delta\frac{\partial \mathcal{P}_k}{\partial x_{jk}}(\tilde{x}_{1k},\ldots,\tilde{x}_{jk},0,\ldots,0)$$

$$= \delta \sum_{k=1}^N \sum_{j \in J} \tilde{x}_{jk}\frac{\partial \mathcal{P}_k}{\partial x_{jk}}(\tilde{x}_{1k},\ldots,\tilde{x}_{jk},0,\ldots,0) \geq \delta \sum_{k=1}^N \mathcal{P}_k(\tilde{x}_{1k},\ldots,\tilde{x}_{nk}) = \delta E_{\text{PD}}.$$

The involved inequality is based on the fact that for any differentiable convex function $f\colon \mathbb{R}^n \to \mathbb{R}$ with $f(0) = 0$ and $x \in \mathbb{R}^n_{\geq 0}$ we have $\sum_{j=1}^n x_j \frac{\partial f}{\partial x_j}(x_1,\ldots,x_j,0,\ldots,0) \geq f(x)$ (see, e.g., [8, Chapter 3]). Together the bounds yield the lemma's statement. ◀

▶ **Lemma 10** (Low-yield Jobs). $g_2(\tilde{\lambda}) \geq \alpha^{-\alpha} \sum_{j \in J_2} v_j$.

**Proof.** Proposition 8(a) together with the fact that $\tilde{\lambda}_j = v_j$ for $j \in J_2$ yields $E_{\tilde{\lambda}}(j) = v_j \frac{\hat{x}_j}{\alpha}$. Applying this to $g_2(\tilde{\lambda})$ we get

$$g_2(\tilde{\lambda}) = \sum_{j \in J_2} (1-\alpha) E_{\tilde{\lambda}}(j) + \sum_{j \in J_2} \tilde{\lambda}_j = \sum_{j \in J_2} \frac{1-\alpha}{\alpha} \hat{x}_j v_j + \sum_{j \in J_2} v_j = \sum_{j \in J_2} \left(1 - \frac{\alpha-1}{\alpha} \hat{x}_j\right) v_j$$

$$\stackrel{\text{Def.}}{\underset{J_2}{\geq}} \sum_{j \in J_2} \left(1 - \frac{\alpha - \alpha^{1-\alpha}}{\alpha}\right) v_j = \alpha^{-\alpha} \sum_{j \in J_2} v_j. \qquad \blacktriangleleft$$

▶ **Lemma 11** (High-yield Jobs). $g_3(\tilde{\lambda}) \geq \frac{1-\alpha}{\alpha^\alpha} \sum_{j \in J_3} E_{\text{PD}}(j) + \alpha^{-\alpha} \sum_{j \in J_3} v_j$ if $\delta \leq \frac{1}{\alpha^{\alpha-1}}$.

**Proof.** We make use of both Proposition 8(a) and Proposition 8(c). First note that the prerequisite $\delta \leq \frac{1}{\alpha^{\alpha-1}}$ together with $\alpha > 1$ and $j \in J_3$ gives us the relation $\delta^{\frac{1}{\alpha-1}} \leq \frac{1}{\alpha} \leq 1 \leq \frac{\alpha - \alpha^{1-\alpha}}{\alpha-1} < \hat{x}_j$. This allows us to apply Proposition 8(c). The second summand of $g_3(\tilde{\lambda})$ is split into two parts, one of which is accounted for by energy invested by PD and the other one by lost value due to unfinished jobs:

$$g_3(\tilde{\lambda}) = \sum_{j \in J_3} (1-\alpha) E_{\tilde{\lambda}}(j) + \sum_{j \in J_3} \tilde{\lambda}_j$$

$$= \sum_{j \in J_3} (1-\alpha) E_{\tilde{\lambda}}(j) + \sum_{j \in J_3} \left(1 - \alpha^{-\alpha}\right) \tilde{\lambda}_j + \sum_{j \in J_3} \alpha^{-\alpha} \tilde{\lambda}_j$$

$$= \sum_{j \in J_3} (1-\alpha) E_{\tilde{\lambda}}(j) + \sum_{j \in J_3} \left(1 - \alpha^{-\alpha}\right) \frac{\alpha E_{\tilde{\lambda}}(j)}{\hat{x}_j} + \sum_{j \in J_3} \alpha^{-\alpha} v_j$$

$$= \sum_{j \in J_3} (1-\alpha) E_{\tilde{\lambda}}(j) \left(1 - \frac{\alpha - \alpha^{1-\alpha}}{(\alpha-1) \hat{x}_j}\right) + \sum_{j \in J_3} \alpha^{-\alpha} v_j$$

$$> \sum_{j \in J_3} (1-\alpha) \delta^{\frac{\alpha}{\alpha-1}} E_{\text{PD}}(j) \left(1 - \frac{\delta^{\frac{1}{\alpha-1}}}{\hat{x}_j}\right)^{-\alpha} \left(1 - \frac{\alpha - \alpha^{1-\alpha}}{(\alpha-1) \hat{x}_j}\right) + \sum_{j \in J_3} \alpha^{-\alpha} v_j$$

$$\geq \sum_{j \in J_3} (1-\alpha) \alpha^{-\alpha} \left(1 - \frac{1}{\alpha \hat{x}_j}\right)^{-\alpha} \left(1 - \frac{1}{\hat{x}_j}\right) E_{\text{PD}}(j) + \sum_{j \in J_3} \alpha^{-\alpha} v_j$$

$$\geq (1-\alpha) \alpha^{-\alpha} \sum_{j \in J_3} E_{\text{PD}}(j) + \sum_{j \in J_3} \alpha^{-\alpha} v_j.$$

The first inequality applies Proposition 8(c), the penultimate inequality the relations deduced from the prerequisite, and the last inequality is the application of Bernoulli's inequality. ◀

## 4.4 Deriving the Tight Competitive Ratio

It remains to derive our final upper bound on the competitive ratio of PD. We do so by combining the bounds from Lemma 9, Lemma 10, and Lemma 11.

▶ **Theorem 3.** *The competitive ratio of algorithm* PD *with the parameter $\delta$ set to $\frac{1}{\alpha^{\alpha-1}}$ is at most $\alpha^\alpha$. Moreover, there is a problem instance for which* PD *is exactly by a factor of $\alpha^\alpha$ worse than an optimal algorithm. That is, our upper bound is optimal.*

**Proof.** If we combine the results from Lemma 9 to Lemma 11 we get

$$\begin{aligned} g(\tilde{\lambda}) &\geq \alpha^{1-\alpha} E_{\text{PD}} + (1-\alpha)\alpha^{-\alpha} \sum_{j \in J_1 \cup J_3} E_{\text{PD}}(j) + \alpha^{-\alpha} \sum_{j \in J_2 \cup J_3} v_j \\ &\geq \alpha^{1-\alpha} E_{\text{PD}} + (1-\alpha)\alpha^{-\alpha} \sum_{j \in J} E_{\text{PD}}(j) + \alpha^{-\alpha} \sum_{j \in J_2 \cup J_3} v_j \\ &\geq \left(\alpha^{1-\alpha} + (1-\alpha)\alpha^{-\alpha}\right) E_{\text{PD}} + \alpha^{-\alpha} \sum_{j \in J_2 \cup J_3} v_j = \alpha^{-\alpha} \text{cost}(\text{PD}). \end{aligned}$$

Now, let OPT denote an optimal schedule for the current problem instance. Moreover, let OPT$'$ denote an optimal solution to the relaxed mathematical program (CP). Obviously, it holds that $\text{cost}(\text{OPT}') \leq \text{cost}(\text{OPT})$. By duality, we know that $g(\tilde{\lambda}) \leq \text{cost}(\text{OPT}')$. By combining these inequalities we can bound PD's competitiveness by

$$\text{cost}(\text{PD}) \leq \alpha^{\alpha} g(\tilde{\lambda}) \leq \alpha^{\alpha} \text{cost}(\text{OPT}') \leq \alpha^{\alpha} \text{cost}(\text{OPT}).$$

For the lower bound, consider a single processor and assume the job values are high enough to ensure that PD finishes all jobs. We create a job instance of $n$ jobs in the same way as done in [3] for the lower bound on OA and AVR. That is, job $j \in J = \{1, 2, \ldots, n\}$ arrives at time $j-1$ and has workload $(n-j+1)^{-1/\alpha}$. All jobs have the same deadline $n$. Now, whenever one of the jobs arrives, PD schedules all remaining jobs at the energy-optimal (i.e., minimal) speed as pool jobs. In other words, it computes a schedule that is optimal for the remaining known work. This is exactly what OA does (hence its name), which means that we get the same lower bound of $\alpha^{\alpha}$ as for OA (cf. [3, Lemma 3.2]). ◀

## 5 Conclusion

We presented a new algorithm and an analysis based on duality theory for scheduling valuable jobs on multiple speed-scalable processors. Using duality theory to approach the analysis of energy-efficient scheduling algorithms was recently proposed by Gupta, Krishnaswamy, and Pruhs [12]. Given that the first formal proof of the original offline algorithm's optimality was achieved by means of duality theory using the KKT conditions [4], it seems that this is a very natural way to approach this kind of problems. However, almost all results for online algorithms in this area use amortized competitiveness arguments similar to the original proof of OA's competitiveness, one of the first and most important online algorithms for energy-efficient scheduling. While this approach proved to be elegant and very powerful, designing suitable potential functions is difficult and needs a quite high amount of experience with the topic. Adapting these potential functions to new model variations and generalizations, or tuning them to narrow the gap to the known lower bounds is non-trivial and remains a challenging task. We think that using well-developed utilities from duality theory for convex programming may prove to be a worthwhile and promising alternative approach. Our results underline this conjecture, not only improving upon known results proved using the classical method but also generalizing them to the important case of multiple processors.